\documentclass{aa}  

\usepackage{graphicx}
\usepackage{natbib}
\usepackage{ulem}
\usepackage{longtable}
\usepackage{subfig}
\usepackage{float}
\usepackage[usenames,dvipsnames]{xcolor}
\usepackage{multirow}
\usepackage{footnote}
\usepackage[bottom]{footmisc}
\usepackage{ulem}

\usepackage{hyperref}
\usepackage{xcolor}
\hypersetup{
  colorlinks   = true, 
  urlcolor     = black, 
  linkcolor    = blue, 
  citecolor   = blue 
}
\usepackage[all]{hypcap}

\usepackage{txfonts}

\newcommand{\teff}{T_\mathrm{eff}}
\newcommand{\logg}{\mathrm{log}\,g}

\begin{document}

    \title{Photospheric properties and fundamental parameters of M dwarfs}

\titlerunning{H-band spectroscopy of M-dwarfs}
 \authorrunning{Rajpurohit et al}

\author{A. S. Rajpurohit \inst{1}, F. Allard \inst{2}, G. D. C. Teixeira \inst{3, 4}, D. Homeier \inst{5}, S. Rajpurohit \inst{6}, O. Mousis \inst{7}}

\institute{Astronomy \& Astrophysics Division, Physical Research Laboratory, Ahmedabad 380009, India
\email{arvindr@prl.res.in}
\and
Univ Lyon, Ens de Lyon, Univ Lyon1, CNRS, Centre de Recherche Astrophysique de Lyon UMR5574, F-69007, Lyon, France
\and
Instituto de Astrof\'{i}sica e Ci\^{e}ncias do Espa\c{c}o, Universidade do Porto, CAUP, Rua das Estrelas, 4150-762 Porto, Portugal
\and
Departamento de F\'{i}sica e Astronomia, Faculdade de Ci\^{e}ncias, Universidade do Porto, Rua Campo Alegre, 4169-007 Porto, Portugal
\and
Zentrum f{\"u}r Astronomie der Universit{\"a}t Heidelberg, Landessternwarte,
K\"{o}nigstuhl 12, 69117 Heidelberg, Germany
\and
Clausthal University of Technology, Institute for Theoretical Physics, Leibnizstr.10, 38678 Clausthal-Zellerfeld, Germany
\and
Aix Marseille Universit\'{e}, CNRS, LAM (Laboratoire d'Astrophysique de Marseille) UMR 7326, 13388, Marseille, France 
}

   \date{Received May 1, 2016; accepted }

 
 \abstract
    {M dwarfs are an important source of information when studying and probing the lower end of the Hertzsprung-Russell (HR) diagram, down to the hydrogen-burning limit. Being the most numerous and oldest stars in the galaxy, they carry fundamental information on its chemical history. The presence of molecules in their atmospheres, along with various condensed species, complicates our understanding of their physical properties and thus makes the determination of their fundamental stellar parameters more challenging and difficult.  
   }
   {The aim of this study is to perform a detailed spectroscopic analysis of the high-resolution H-band spectra of M dwarfs in order to determine their fundamental stellar parameters and to validate atmospheric models. The present study will also help us to understand various processes, including dust formation and depletion of metals onto dust grains in M dwarf atmospheres. The high spectral resolution also provides a unique opportunity to constrain other chemical and physical processes that occur in a cool atmosphere.
   }  
   {The high-resolution APOGEE spectra of M dwarfs, covering the entire H-band, provide a unique opportunity to measure their fundamental parameters. We have performed a detailed spectral synthesis by comparing these high-resolution H-band spectra to that of the most recent BT-settl model and have obtained fundamental parameters such as  effective temperature,
surface gravity, and metallicity ($\teff$, $\logg$ and [Fe/H], respectively).
   }
   {We have determined $\teff$, $\logg$ and [Fe/H] for 45 M dwarfs using high-resolution H-band spectra. The derived $\teff$ for the sample ranges from 3100 to 3900 K, values of $\logg$  lie in the range 4.5 $\le$ $\logg$
$\le$ 5.5, and the resulting metallicities lie in the range -0.5 $\le$ [Fe/H] $\le$ +0.5. We have explored systematic differences between effective temperature and metallicity calibrations with other studies using the same  sample of M dwarfs. We have also shown that the stellar parameters determined using the BT-Settl model are more accurate and reliable compared to other comparative studies using alternative models.
   }
   {}

   \keywords{Stars: low-mass -- M dwarfs --Stellar atmosphere -- fundamental parameters -- atmospheres-- late type}

   \maketitle
%

\section{Introduction}
The lower end of the Hertzsprung-Russell (HR) diagram has proven extremely useful in the last few decades as most of the very-low-mass stars (VLM) in the Galaxy are located in that region. Seventy percent of the Galactic stellar population \citep{Bochanski2010} consists of these VLMs, in particular M dwarfs, and they contribute approximately 40$\%$ of the total stellar mass budget of the Galaxy \citep{Gould1996, Mera1996,Henry1998}. Depending on its metallicity, the mass of any particular M dwarf ranges from 0.6 $M_{\odot}$ to the hydrogen-burning limit of about 0.075 $M_{\odot}$ \citep{Chabrier2000}. M dwarf populations show great diversity, in that one can find young metal-rich M dwarfs in open clusters, whereas galactic halos \citep{Green1994} and the globular clusters \citep{Cool1996,Renzini1996} are known to host metal-poor M dwarfs that are  billions of years old. Therefore, M dwarfs are one of the most important stellar components of the Galaxy, carrying  fundamental information on the Galaxy's structure, formation, and  chemical history. Recently, brown dwarfs and super earths were found around M dwarfs  \citep{Bonfils2012,Anglada2016,Gillon2017} which makes them an important laboratory to study and understand their formation.

Despite the large number of M dwarfs in the Galaxy, a homogenous sample, in terms of age and metallicity, is very difficult to obtain, as high-resolution images and good S/N spectra are rare because of their intrinsics faintness. Moreover, the non-existence of true continuum makes it difficult or impossible to isolate different spectral diagnostics and to disentangle the effective temperature, surface gravity, and metallicity ($\teff$, $\logg$, and [Fe/H]). The presence of diatomic and triatomic molecules along with dust in M dwarf atmospheres, as we go from early to late M dwarfs, makes access to the spectral continuum nearly impossible both in the optical and in the near-infrared (NIR). Nevertheless, because of their cool temperature and low metal content, M dwarfs provide the best laboratory to study the dust and cloud formation as well as radiative transfer in their atmosphere.

As the $\teff$ of M dwarfs decreases from early to late M dwarfs, the optical and NIR spectra of M dwarfs indicate large numbers of diatomic (SiH, CaH, TiO, VO, CrH, FeH, OH, CO) and triatomic (CaOH, H$_2$O) molecules. The Rayleigh-Jeans branch of M dwarfs spectral energy distribution (SED) in Infra-Red (IR) (\textgreater 1.3 $\mu$m) is dominated by the H$_2$O and CO molecular absorption bands, whereas in the corresponding optical part (\textgreater 0.4 $\mu$m) to near-IR  (\textless 1.3 $\mu$m) their SED is governed by TiO, VO and metal hydrides. Due to the presence of the these complex and crowded band structures, access to true continuum is not possible, and thus a pseudo-continuum is created, which usually shows the strongest and often resonant atomic lines at lower resolution \citep{Allard1990,Allard1995}. In cooler M dwarfs with spectral type M6 or later, the outermost temperatures of their atmospheres are cool enough to form dust and clouds \citep[see e.g.,][]{Tsuji1996a,Tsuji1996b,Allard1997,Ruiz1997,Allard1998}. These various physical and chemical processes complicate the understanding of their cool atmospheres, thus making determination of their stellar properties even more difficult.

The proper classification of M dwarf spectra requires the comparison of a grid of synthetic spectra with observations. These comparisons can thus be used  to derive M dwarfs' fundamental parameters. Such comparisons also help to disentangle and quantify basic physical properties and fundamental parameters such as $\teff$, $\logg$, and [Fe/H]. Thus far,  $\teff$, $\logg$, and [Fe/H]) have not been determined for M dwarfs with great accuracy. Different groups have used various traditional techniques to estimate the $\teff$ of M dwarfs based on broadband photometry and black-body approximations. These relatively old, traditional techniques are not as reliable because the true continua of cool M-dwarfs is embedded in complex and broad molecular absorptions. Furthermore, in M dwarf atmospheres, the complexity increases significantly as dust and cloud formation occurs with decreasing $\teff$. In the optical part of their SEDs, this can be seen as the weakening of TiO-, VO-, CaH-, and CaOH-band opacities by dust Rayleigh scattering, whereas as in the infrared (IR) region, the weakening of water bands occurs due to the greenhouse effect \citep{Allard2001} or dust back-warming.

Models of the atmospheres of cool, low-mass stars and substellar objects have been the subject of tremendous development in recent decades \citep{Brott2005,Helling2008a,Allard2012,Allard2013}. Because of this advancement, a number of studies is being carried out to derive the accurate stellar parameters of very-low-mass stars and brown dwarfs using both optical and near-infrared (NIR) observations \citep{Burgasser2006,Bayo2014,Bayo2017,Rajpurohit2012a,Rajpurohit2013,Rajpurohit2014,Rajpurohit2016}. \cite{Bayo2017} and  \cite{Bayo2014} show the importance of consistent fundamental parameters by estimating their atmospheric parameters from optical and in the NIR with low-resolution spectra and photometry of M dwarfs, simultaneously. Through revised solar abundances by \cite{Asplund2009} and \cite{Caffau2011}, and by incorporating updated atomic and molecular line opacities which govern the SED of M dwarfs, atmospheric models such as the BT-Settl \citep{Allard2013} have seen major improvements in modeling various complex molecular absorption bands. These updated models now also include dust and cloud formation \citep{Allard2013,Baraffe2015}, which is important for cool M-dwarfs and metal-poor M subdwarfs (sdM), and thus yield promising results which explain the stellar-to-substellar transition and confirm the work of \cite{Rajpurohit2012a}.

In comparison to our Sun, the determination of atmospheric parameters for M dwarfs is very different and challenging. Stellar parameters, such $\teff$ , of M dwarfs remain model-dependent to some extent. There have been many attempts to derive the $\teff$ scale of M dwarfs with respect to constant age and metallicity. Due to the lack of very reliable model atmospheres in the past, \cite{Bessell1991} used indirect methods to derive the $\teff$ scale of M dwarfs  based on black-body fitting to the NIR bands, whereas \cite{Wing1979} and \cite{Veeder1974} fitted much cooler black body to the optical. \cite{Tsuji1996a} and \cite{Casagrande2008} provided a good $\teff$ determination using an infrared flux method (IRFM) for dwarfs including M dwarfs. The M dwarfs in the Rayleigh Jeans tail (mostly red-wards of 2.5 $\mu$m) carry little flux compared to black body, thus the IRFM method tends to underestimate $\teff$ for M dwarfs. \cite{Boyajian2012} used another approach which is based on interferometrically determined radii and bolometric fluxes from photometry to calculate the $\teff$ for nearby K and M dwarfs,  whereas \cite{Mann2015} determined the radius and mass by combining the empirical mass-luminosity relationships with evolutionary models, which in turn depend on the $\teff$ and metallicity.

\begin{figure*}[!thbp]
\centering
\includegraphics[width=15.5cm,height=15cm,angle=-90]{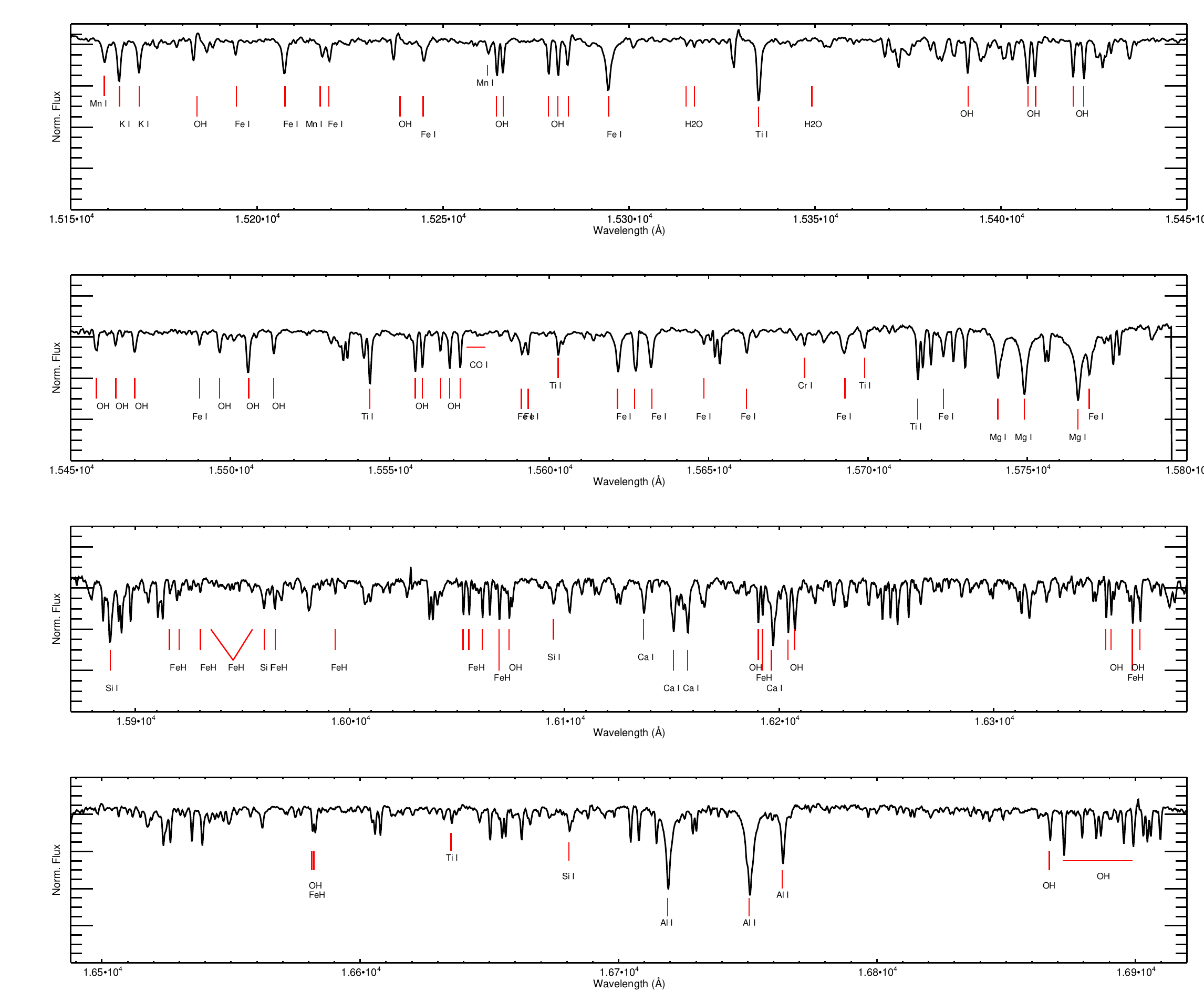}
\caption{APOGEE spectra of 2M11091225-0436249 (M0.5). The main spectral features can be seen,  including atomic lines such as Fe I, Ca I, Na I, K I, Si I, Mg I, Al II, along with some hydride bands such as those of FeH and OH. We used mainly \cite{Souto2017} for the spectral feature recognition.}
\label{Fig1}
\end{figure*}

Recently, \cite{Rajpurohit2013} determined the $\teff$ of nearby bright M dwarfs from the low-resolution spectra observed in the visible wavelength using the updated BT-Settl model atmosphere.  Their study shows that these updated models can now reproduce the slope of their SED very well, unlike previous studies by \cite{Leggett1996,Leggett1998,Leggett2000,Leggett2001} using previous versions of these models, which were using incomplete opacities and other inaccuracies. The $\logg$ of M dwarfs can be determined with the help of high-resolution spectra \citep{passegger2016, Rajpurohit2016}. \cite{passegger2016, Rajpurohit2016} used gravity-sensitive features such as Na I, K I and Ca I lines to determine the surface gravity in the optical.  Other authors used interferometry  to determine the angular diameter of the M dwarfs, together with mass-luminosity relations to derive the mass and $\logg$ (e.g. \cite{Segransan2003}). 

A proper metallicity calibration for M-dwarfs is essential to determine the planet star metallicity relation, which for FGK stars tends towards the super solar metallicities. The metallicity determination of M dwarfs can be done in two ways: photometric- and spectroscopic-based methods which are limited to the moderate-resolution spectra in the visible \citep{Woolf2006,Woolf2009}, and in the infrared \citep{Mann2013,Mann2014,Terrien2012,Rojas2010,Newton2014}. The former techniques use M dwarf photometry in the visible and infrared bands to create [Fe/H] calibrations \citep{Bonfils2005,Johnson2009,Schlaufman2010}, while the latter ones rely on low- to high-resolution spectra to measure indices and lines in order to establish spectroscopic calibrations or compare them to synthetic spectra, made from M dwarf atmospheric models \citep{Valenti1998,Bean2006b,Bean2006a,Lindgren2017}. Recently \cite{Souto2017} presented the first detailed near-IR chemical abundance analysis observed by SDSS-IV-Apache Point Observatory Galactic Evolution Experiment (APOGEE, \cite{Majewski2015}). The $\teff$ values adopted in this study were derived from the photometric calibrations for M dwarfs by \cite{Mann2015} for the V-J and R-J colors.

In this paper, we take advantage of the updated BT-Settl model grid and high-resolution H-band spectra to determine the atmospheric parameters ($\teff$, ${\logg}$ and [Fe/H]) of 45 M dwarfs. In Section \ref{obs}, we briefly describe the observations and some aspects of data reduction. In Section \ref{models}, we describe the BT-Settl model atmosphere used in this study. Section \ref{results} presents the results and describes the comparison with models and determination of stellar parameters. A Summary and Discussion are presented in Section \ref{discussion}.

\section{Observational data and sample selection}
\label{obs}
Regarding the details of the APOGEE survey along with data reduction the reader is referred to \citep{Majewski2015,Wilson2010,Wilson2012}. The details of APOGEE M dwarfs ancillary project along with target selection and data reduction are described in \cite{Deshpande2013} and \cite{Nidever2015}. We obtained spectra of 45 M dwarfs from \cite{Deshpande2013} M dwarfs ancillary project using SDSS-III Data release 12 \citep{Alam2015}. The spectral type and photometry are compiled using Simbad and Vizier catalog access through Centre de Donnees astronomiques de Strasbourg and are given in Table \ref{Table1}. 

\begin{figure*}[!thbp]
        \centering
        \includegraphics[width=11.5cm,height=15cm,angle=-90]{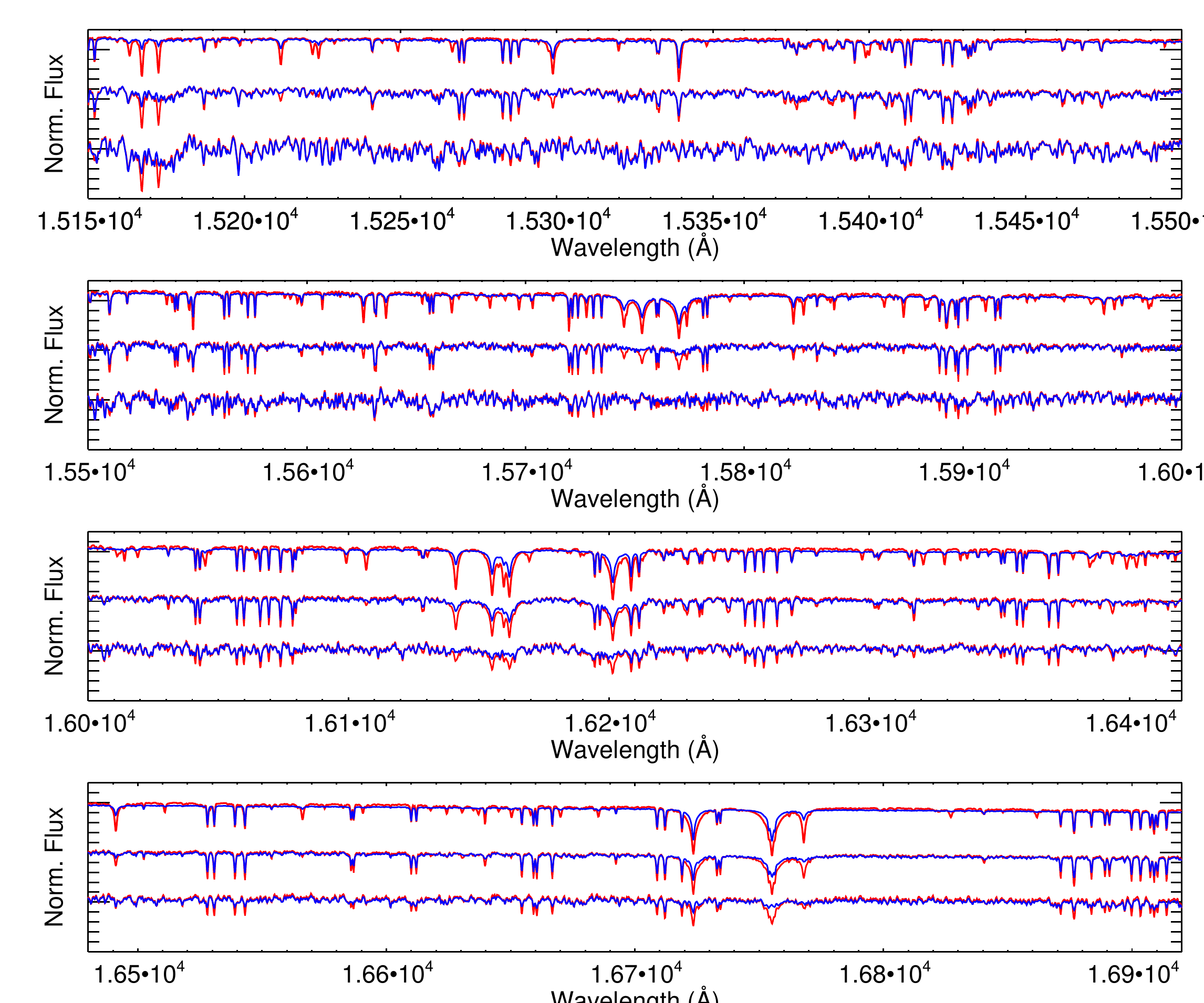}
\caption{BT-Settl synthetic spectra from 4000~K to 3000~K at a step of 500~K (top to bottom in each panel) of H-band computed  with PHOENIX radiative transfer code. The red and blue lines represent the synthetic spectra at [Fe/H] = +0.5 (red) and -0.5 (blue) for  $\teff$ = 4000K , 3500K and 3000K at constant log g of 5.5.}
\label{Fig2}
\end{figure*}

\begin{table*}[!thbp]
        \centering
        \caption{NIR photometry for our sample is taken from 2MASS along with their coordinates and spectral types.}
        \begin{tabular}{lllllllll}
                \hline\\
                2MASS ID& J &H &K$_s$ &$\alpha$ &$\delta$ &SpT\\
                2MXXXXX&&&&&&\\
                \hline\\     
  00131578+6919372& 08.55$\pm$0.024 & 07.98$\pm$0.02  &07.74$\pm$0.02 &03.315773     &   69.327003&M3.0\\
  00321574+5429027& 09.38$\pm$0.022 & 08.82$\pm$0.01  &08.57$\pm$0.01 &08.065590    &    54.4841&M4.5\\  
  00350487+5953079& 11.03$\pm$0.022 & 10.40$\pm$0.02  &10.16$\pm$0.02  &08.77032&    59.885548&M4.3\\
  01195227+8409327& 09.85 $\pm$0.026 & 09.31$\pm$0.03 &09.02$\pm$0.02  &19.967825&  84.159111&M5.0\\
  02085359+4926565& 08.42$\pm$0.023 & 07.81$\pm$0.01  &07.58$\pm$0.02  &32.223315 & 49.449055&M4.0\\
  03152943+5751330& 11.12$\pm$0.024 & 10.53$\pm$0.03  &10.27$\pm$0.01  &48.872662&  57.85918&M3.5 \\
  03305473+7041145&  09.48$\pm$0.018 &08.93$\pm$0.01  &08.67$\pm$0.01   &52.728069&  70.687378&M3.5\\
  03425325+2326495& 10.20$\pm$0.022 & 09.54$\pm$0.02  &09.31$\pm$0.02   &55.721897&  23.447109&M4.0\\
  04063732+7916012& 10.03$\pm$0.027 & 09.48$\pm$0.02  &09.19$\pm$0.02   &61.655503&  79.267006&M4.5\\
  04125880+5236421& 08.77$\pm$0.032 & 08.24$\pm$0.03  &07.91 $\pm$0.01  &63.245023&  52.611698&M4.0\\
  05011802+2237015& 10.16$\pm$0.020 & 09.59$\pm$0.02  &09.23$\pm$0.01   &75.325112&  22.617104&M5.0\\
  05030563+2122362& 09.75$\pm$0.021 & 09.16$\pm$0.02  &08.88$\pm$0.01   &75.773472&  21.376726&M5.0\\
  05210188+3425119& 11.87$\pm$0.021  & 11.31$\pm$0.01  &11.02$\pm$0.01   &80.257859&  34.419991&M5.0\\
  05470907-0512106& 10.03$\pm$0.024  & 09.51$\pm$0.02  &09.17$\pm$0.01   &86.787800&    -5.202969&M4.5\\
  06115599+3325505& 10.16$\pm$0.019 & 09.59$\pm$0.02  &09.34$\pm$0.02   &92.983296&  33.430714&M3.5\\
  06320207+3431132& 10.69$\pm$0.021 & 10.14$\pm$0.01  &09.86 $\pm$0.01  &98.008631&  34.520336&M4.0\\
  07140394+3702459& 11.97$\pm$0.023 & 11.25 $\pm$0.03 &10.83$\pm$0.01& 108.516439 &37.046108&M8.0\\
  08501918+1056436& 11.28$\pm$0.023 & 10.67 $\pm$0.02 &10.40$\pm$0.02& 132.579937& 10.945469&M5.0\\
  09301445+2630250& 08.86$\pm$0.020 & 08.28 $\pm$0.02 &08.02$\pm$0.02 &142.560229& 26.506958&M3.0\\
  10162955+0318375& 10.85$\pm$0.023 & 10.26 $\pm$0.02 &10.00$\pm$0.02 &154.123134& 3.310419&M4.1\\
  11005043+1204108& 10.67$\pm$0.024 & 10.11$\pm$0.02  & 09.78$\pm$0.02& 165.210134& 12.069667&M5.0\\
  11054316+1014093& 08.64$\pm$0.021 &  08.04$\pm$0.05 & 07.79$\pm$0.02& 166.429854& 10.235927&M3.0\\
  11091225-0436249&  08.20$\pm$0.026 &  07.59 $\pm$0.04&   07.33$\pm$0.02  &167.30107 & -4.606939&M0.5\\
  11474074+0015201& 08.99$\pm$0.035 &  08.39$\pm$ 0.04& 08.09$\pm$0.02& 176.919765& 0.255604&M4.0\\
  12045611+1728119& 09.79$\pm$0.021 &  09.18 $\pm$ 0.02& 08.96$\pm$0.02 & 181.233799& 17.469975&M3.5\\
  12232063+2529441& 10.82$\pm$0.019 & 10.23 $\pm$0.01&  09.98$\pm$0.01& 185.83597&  25.495592&M3.7\\
  12265737+2700536& 10.19$\pm$0.024 &  09.60$\pm$ 0.02& 09.32$\pm$0.02& 186.739043& 27.014906&M4.5\\
  13085059+1622039& 09.26$\pm$0.022 &  08.65$\pm$ 0.02&  08.41$\pm$0.01& 197.210793& 16.36775&M3.0\\
  13345147+3746195& 09.71$\pm$0.02   &  09.14$\pm$ 0.02& 08.88$\pm$0.01& 203.714472& 37.772106&M3.5\\
  13451104+2852012& 09.88$\pm$0.022 &  09.31$\pm$ 0.02& 09.05$\pm$0.01& 206.296026& 28.867016&M3.4\\
  14592508+3618321& 10.25$\pm$0.018 &  09.64$\pm$ 0.01& 09.37$\pm$0.01& 224.854502& 36.308922&M3.5\\
  16370146+3535456& 11.13$\pm$0.022 &  10.54$\pm$ 0.02 &10.24$\pm$0.01& 249.256085 &35.596016&M6.0\\
  18451027+0620158& 07.65$\pm$0.019 &  07.04$\pm$ 0.02& 06.80$\pm$0.02& 281.292808 &6.337733&M1.0\\
  18523373+4538317& 10.49$\pm$0.020 &  09.93$\pm$ 0.01& 09.67$\pm$0.01& 283.140551& 45.642147&M5.0\\
  18562628+4622532& 09.59$\pm$0.021 &  09.01$\pm$ 0.01& 08.71$\pm$0.01& 284.109528& 46.381451&M4.0\\
  19051739+4507161& 09.85$\pm$0.021 &  09.30$\pm$ 0.01& 09.02$\pm$0.01 &286.322483 &45.121147&M4.0\\
  19071270+4416070& 10.44$\pm$0.020 &  09.85 $\pm$0.01& 09.55$\pm$0.01& 286.802929 &44.268635&M4.5\\
  19081576+2635054& 10.36$\pm$0.024 &  09.76$\pm$0.03& 09.47$\pm$0.02& 287.065699 &26.584858&M5.0\\
  19084251+2733453& 09.75$\pm$0.026 &  09.23$\pm$0.03& 08.95$\pm$0.01 & 287.177127 &27.562593&M4.3\\
  19321796+4747027& 11.51$\pm$0.020 &  10.93$\pm$0.01& 10.63 $\pm$0.02& 293.074865 &47.78409&M5.0\\
  19333940+3931372& 08.12$\pm$0.020 &  07.56$\pm$0.02& 07.33 $\pm$0.01 & 293.414198& 39.527016&M2.0\\
  19430726+4518089& 11.33$\pm$0.023 &  10.75$\pm$0.02& 10.38$\pm$0.01 & 295.780281 &45.302483&M5.5\\
  19443810+4720294& 11.81$\pm$0.021 &  11.28$\pm$0.01& 11.00$\pm$0.01  & 296.158759& 47.341515&M4.5\\
  19510930+4628598& 08.58$\pm$0.023 &  08.04$\pm$0.02& 07.77$\pm$0.01 &297.788774 &46.483295&M4.0\\
  21105881+4657325& 09.87$\pm$0.022 &  09.26$\pm$0.01& 09.05$\pm$0.01 &317.745051& 46.959034&M3.5\\
                \hline\\
                \label{Table1}
        \end{tabular}
\end{table*}

The presence of broad and complex molecular absorption in H-band makes this region one of the most difficult wavelength regimes for identifying various weak atomic absorption features in the spectra of M dwarfs. The dominant NIR features are due to photospheric absorption by H$_2$O, FeH, CO, OH, and neutral metals. The absorption lines of neutral metals, as well as the bands of H$_2$O and CO, become stronger with decreasing $\teff$. In the optical region, M dwarfs show strong features relative to the strength of the TiO and VO molecular bands. However, in the NIR regime, the dominant molecular features are due to H$_2$O. Also the single metal species such as FeH will not show the same level of decrease as the double metal TiO. The effect of collisional induced absorption (CIA) by H$_2$ on atomic spectral lines such as those of Fe I, Ca I, Na I, K I, Si I, Mg I, Al I, along with the strengthening of hydride bands such as those on FeH can be seen in their H-band spectra (Fig~\ref{Fig1}). As expected from the high $\logg$ of M dwarfs, the atomic features such as Ca I, Na I, and K I are massively pressure broadened. The OH and FeH produce more diffuse absorption NIR, unlike TiO and VO which produce distinctive band heads in the optical. The significant opacity source in the H band of M dwarfs is mainly FeH but its relative strength decreases and becomes saturated with decreasing temperature. In general one can see various prominent atomic lines such as Ca I, Na I, K I, Si I, Mg I, and Al I throughout all the observed spectra. However, it is difficult to identify and measure the intensities of these atomic lines in the region where strong molecular absorption features are present.

The Ca I lines at 1.6136~$\mu$m, 1.6150~$\mu$m and 1.6157~$\mu$m, K I lines at 1.5163~$\mu$m and 156168 $\mu$m, Mg I lines at 1.5740~$\mu$m, 1.5748~$\mu$m, and 1.5765~$\mu$m, and Al I lines at 1.6718, 1.6750 and 1.6763 $\mu$m  can bee seen in all the observed spectra. These atomic lines become broadened as one goes from early to later M dwarfs. As these atomic lines are so broad, their equivalent widths (EW)  are of several angstroms. The strengths of these atomic features depend on various stellar parameters such as $\teff$, [Fe/H], and luminosity. These atomic lines which are relatively free from any blends, and which are not contaminated by telluric lines, are ideal features for studying their sensitivity to various stellar parameters.

\begin{figure*}[!thbp]
        \centering
        \includegraphics[width=8.5cm,height=18cm,angle=-90]{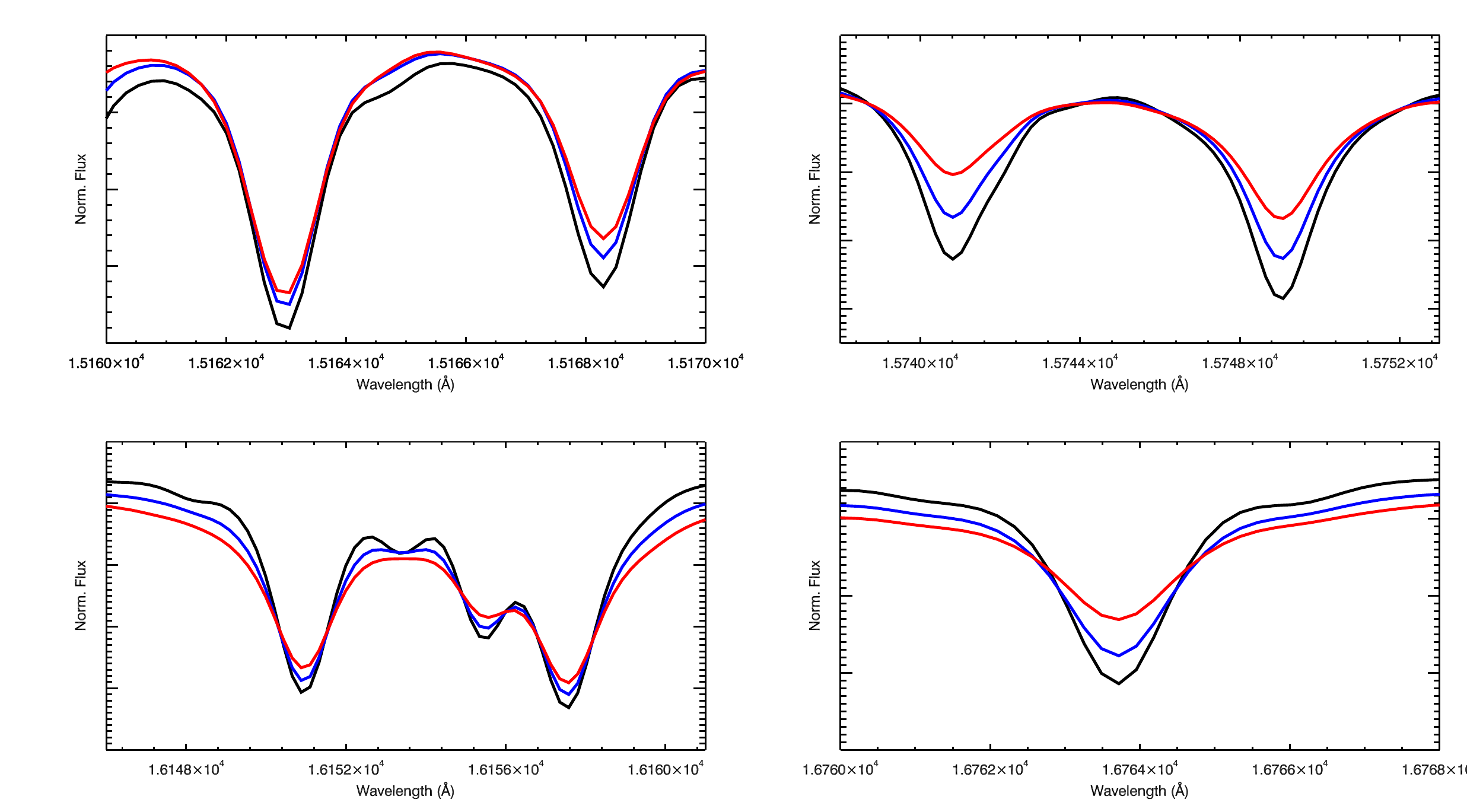}
        \caption{The effect of $\logg$ on the K I (top left), Mg I (top right), Ca I (bottom left) and Al I (bottom right) is clearly visible in BT-Settl synthetic spectra at $\teff$ = 3500 K and  $\logg$ = 4.5 (black), 5.0 (blue),and 5.5 (red).}
   \label{Fig3}
\end{figure*}

\begin{figure*}[!thbp]
        \centering
        \includegraphics[width=11.5cm,height=15cm,angle=-90]{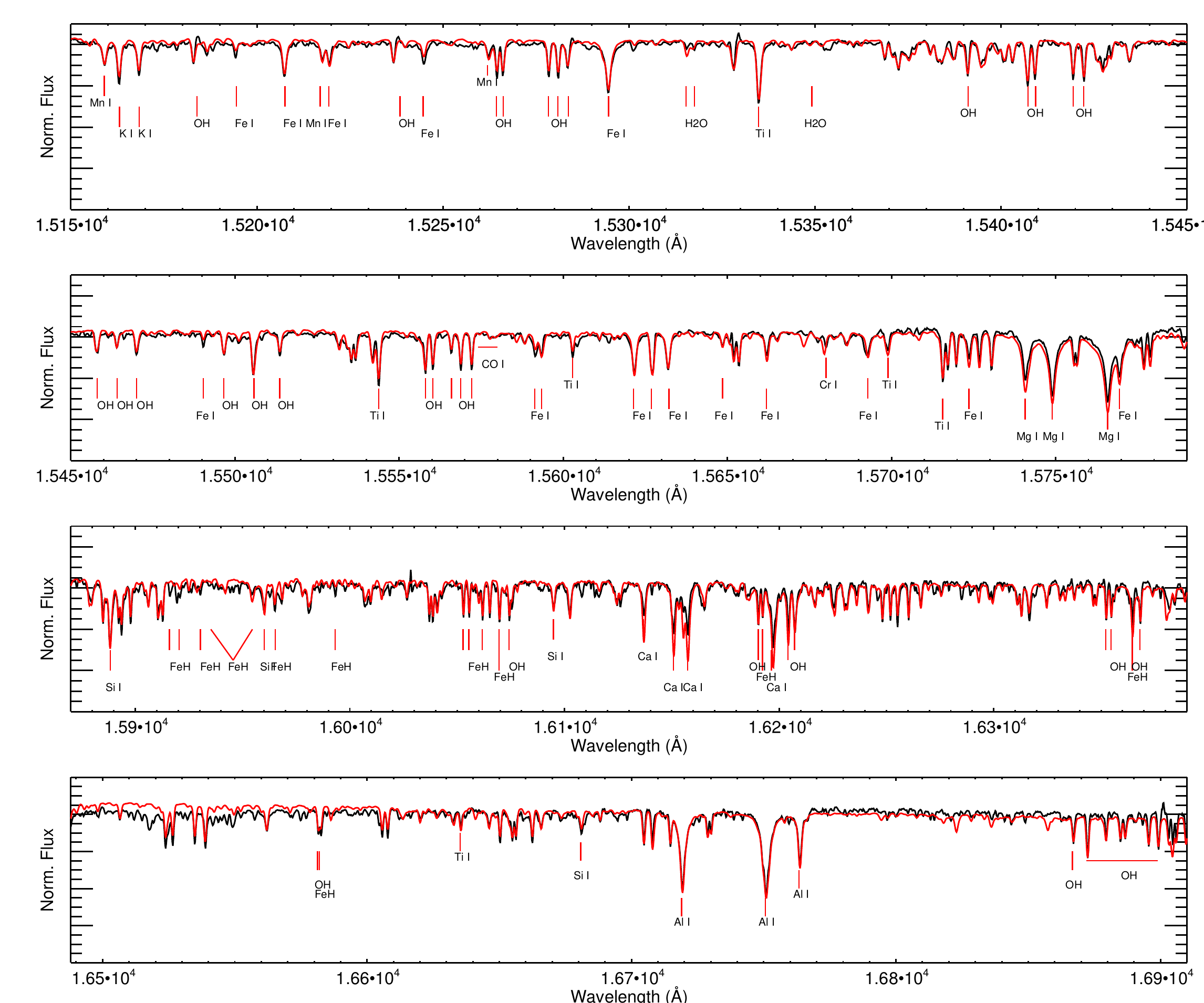}
        \caption{APOGEE spectra of 2M11091225-0436249 (black) of spectral type M0.5 is compared with the best-fit BT-Settl (red). The best fit value for $\teff$, $\logg$ and [Fe/H] is 3900/4.5/-0.3.}
        \label{Fig4}
\end{figure*}

\begin{figure*}[!thbp]
        \centering
        \includegraphics[width=11.5cm,height=15cm,angle=-90]{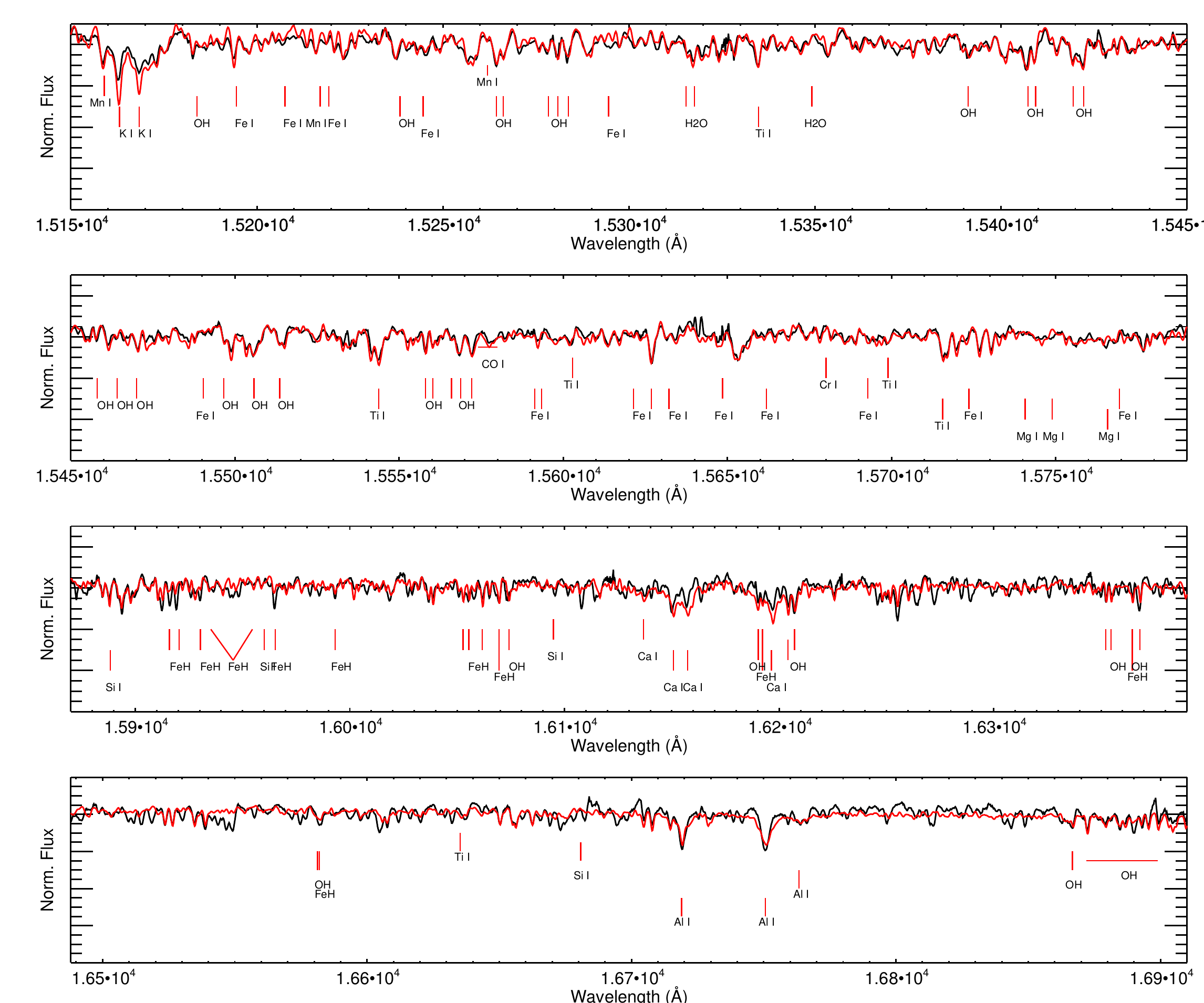}
        \caption{APOGEE spectra of 2M08501918+1056436 (black) of spectral type M5.0 is compared with the best-fit BT-Settl (red) synthetic spectra. The best fit $\teff$, $\logg$ and [Fe/H] is 3100/5.5/-0.0.}
        \label{Fig5}
\end{figure*}

\section{Models and synthetic spectra}
\label{models}

BT-Settl model atmosphere published by \cite{Allard2012,Allard2013} is used in this current study. Their computation of these models is performed with the PHOENIX radiative transfer code \citep{Hauschildt1997,Allard2001} by assuming the hydrostatic and chemical equilibrium, convection using the mixing-length theory and a sampling treatment of the opacities. The grid of BT-Settl models extends from $\teff$ 300 to 7000 K in steps of 100 K, log g = 2.5 to 5.5 in steps of 0.5, and [M/H] = -2.5 to +0.5 in steps of 0.5 dex , accounting for alpha-enhancement and the latest solar abundances by \cite{Asplund2009} and \cite{Caffau2011}. The adopted [$\alpha$/Fe] =-0.4 x [Fe/H] for -1 $\le$ [Fe/H] $\le$ 0 and [$\alpha$/Fe] = +0.4 for all lower and +0.0 for supersolar metallicities, thus setting the "knee" of the alpha-enrichment relation to an average disk population value. These different $\alpha$ enhancements are mainly for the thin disc and thick disc \citep{Edvardsson1993,Gratton1996,Fuhrmann1998,Adibekyan2013}. At a step of 0.1 dex in  $\logg$ and metallicity we have interpolated the grid.  The effect of metallicity and $\teff$ on various atomic and molecular features can bee seen in Figure~\ref{Fig2} with varying $\teff$ from 4000~K (top) to 3000 K (bottom) with a step of 500~K and [Fe/H] = +0.5 (red) and -0.5 (blue) at constant $\logg$ of 5.0 in each panel. As found in previous studies \citep{Leggett1998,Leggett2000}, $\logg$ has a relatively small influence on the SED of M dwarfs. But the significant effect of $\logg$ can been seen at high resolution on various atomic line profiles and also on various band systems, whereas metallicity has a large effect on the spectra. We have shown such effects in Figure~\ref{Fig2} where one can see that with decreasing $\teff$ , various atomic features start vanishing and molecular bands begin to dominate; in particular OH and FeH.

\section{Results}
\label{results}
\subsection{Comparison with models and determination of stellar parameters}

Spectral synthesis using synthetic spectra requires various parameters such as $\teff$, $\logg$, and [Fe/H] and keeping the Sun as a reference. We followed the same procedure as used in \cite{Rajpurohit2014,Rajpurohit2016} to determine $\teff$, $\logg,$ and [Fe/H] using spectroscopic information in H-band. The typical $\logg$ of M dwarfs is approximately 5.0$\pm$0.2, except for the latest-ype M dwarfs \citep{Gizis1997} and \cite{Casagrande2008}; we therefore use models with $\logg$ = 4.5 - 5.5 for our analysis. To determine the stellar parameters of M dwarfs in our sample, we performed a $\chi^2$ minimization using spectral synthesis employing the new BT-Settl model atmospheres across the entire wavelength range of the observed spectra. No weights are applied in our calculation for different parameters. The synthetic spectral fitting is performed using the following steps: In the first step the synthetic spectra are convolved with an isotropic Gaussian profile with measured instrumental resolution which is then interpolated at each wavelength point of the observed spectra. In the following step we compared the observed spectra with that of the entire grid of models by taking the difference between the flux values of the observed and synthetic spectra at every wavelength point. Then, the sum of the squares of these differences is obtained for each model in the grid, and the best model for each object is selected. We retain the best-match values of $\teff$, $\logg$ and [Fe/H] as first guess values on these three parameters. This step of synthetic spectral fitting is performed on the set of models which have not been interpolated to a finer grid in $\logg$ and [Fe/H]. The comparison is made using a subsample of the model atmosphere grid covering the range of 3000 K $\leq$ $\teff$  $\leq$ 4000~K in steps of 100~K, -0.5 $\leq$ [Fe/H] $\leq$ 0.5 in steps of 0.5 dex, and 4.0 $\leq$ log\,$g$ $\leq$ 5.5 in steps of 0.5 dex. During this step, we keep all the three parameters ($\teff$, $\logg$ and [Fe/H]) free. We excluded the spectral regions from 1.580 $ \mu$m to 1.586 $\mu$m and from 1.642 $\mu$m to 1.649 $\mu$m because of the gap in blue to green and green to red arms of APOGEE. 

In the second step, the parameters obtained for each object of our sample from the first step are used as an initial guess value and interpolation is done at a step of 0.1 dex in $\logg$ and [Fe/H]. Finally, every model of the grid covering the range of 3000 K $\leq$ $\teff$  $\leq$ 4000~K in steps of 100~K, -0.5 $\leq$ [Fe/H] $\leq$ 0.5 in steps of 0.1 dex, and 4.0 $\leq$ log\,$g$ $\leq$ 5.5 in steps of 0.1 dex are again compared to the observed spectrum at each wavelength point, and the $\chi^2$ is calculated to determine the global minima. We retain models that give the lowest $\chi^2$ values as the best fit parameters ($\teff$, {$\logg$} and [Fe/H]) which are showed in Table \ref{Table2}. In the end, the best models are finally inspected visually by comparing them with the corresponding observed spectra. The uncertainties in Table 2 are based on standard deviation of the derived stellar parameters by accepting 1 $\sigma$ variations from the minimum $\chi^2$ which in all cases is calculated using constant $\chi^2$ boundaries and is based on the $\chi^2$ statistic.

We have also checked the behavior of synthetic spectra by visual inspection, looking at the shapes of various atomic species such as Fe I, Ca I, Na I, K I, Si I, Mg I, Al I  and some molecular species such as OH, CO, and FeH \citep[for details of the line list, see][]{Souto2017}. The OH-bands around 1.540 to 1.545~$\mu$m, 1.635~$\mu$m to 1.636~$\mu$m and 1.686~$\mu$m to 1.689~$\mu$m are insensitive to variations of 0.5 dex in $\logg$ but are rather highly sensitive to $\teff$. However, at a given $\logg$ they shows huge variation over a change of only 100 K in $\teff$. We have conformed our $\logg$ by looking at the width of gravity-sensitive features such as the K I (1.5163~$\mu$m and 1.5168~$\mu$m) , Ca I (1.6136~$\mu$m, 1.6150~$\mu$m, and 1.6157~$\mu$m), Al (1.6718~$\mu$m, 1.6750~$\mu$m, and 1.6763~$\mu$m), and Mg I (1.5740~$\mu$m, 1.5748~$\mu$m, and 1.5765~$\mu$m) along with the relative strength of metal hydride bands such as FeH. These features are particularly useful gravity discriminants for M dwarfs and sdM. The overall line strength increases with gravity because of high pressure mainly by H$_2$, He, and H I collisions and due to higher electron pressure on alkali lines (see Fig.~\ref{Fig3}). The effect of $\logg$ can be seen on width of the damping wings which, in addition, increases \citep{Reiners2005,Reiners2016a}. The effect of metallicity can also be seen on various atomic features where the molecular absorption is less and these atomic features appear clearly. The synthetic spectrum reproduces line profiles of various atomic lines such as Ti, Fe I, Ca I, Mg I, Si I, Mn I and Al I relatively well. The systematic errors are not eliminated, which arise due to missing or incomplete opacity sources (e.g., FeH-bands and OH and CO bands) are not eliminated \citep{Baraffe2015} and the derived uncertainties are within the derived values of $\teff$, {$\logg,$} and [Fe/H]

\begin{figure}[!thbp]
        \centering
        
        \includegraphics[width=8.0cm,height=4.0cm]{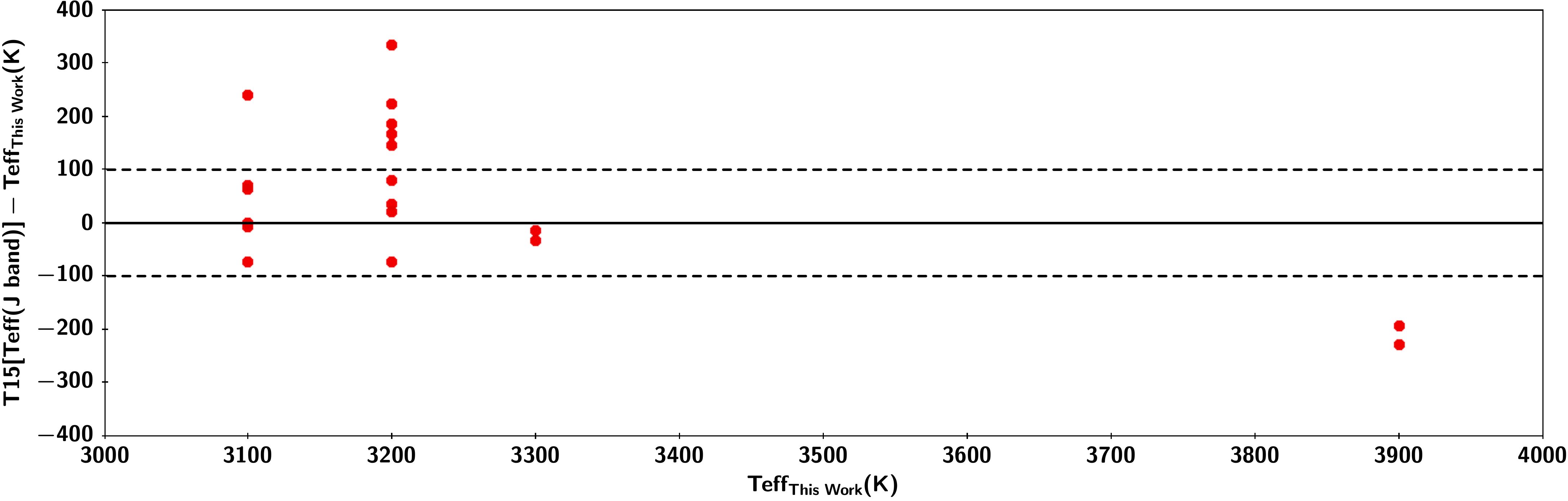}
        
        \hspace{0mm}
        
          \includegraphics[width=8.0cm,height=4.0cm]{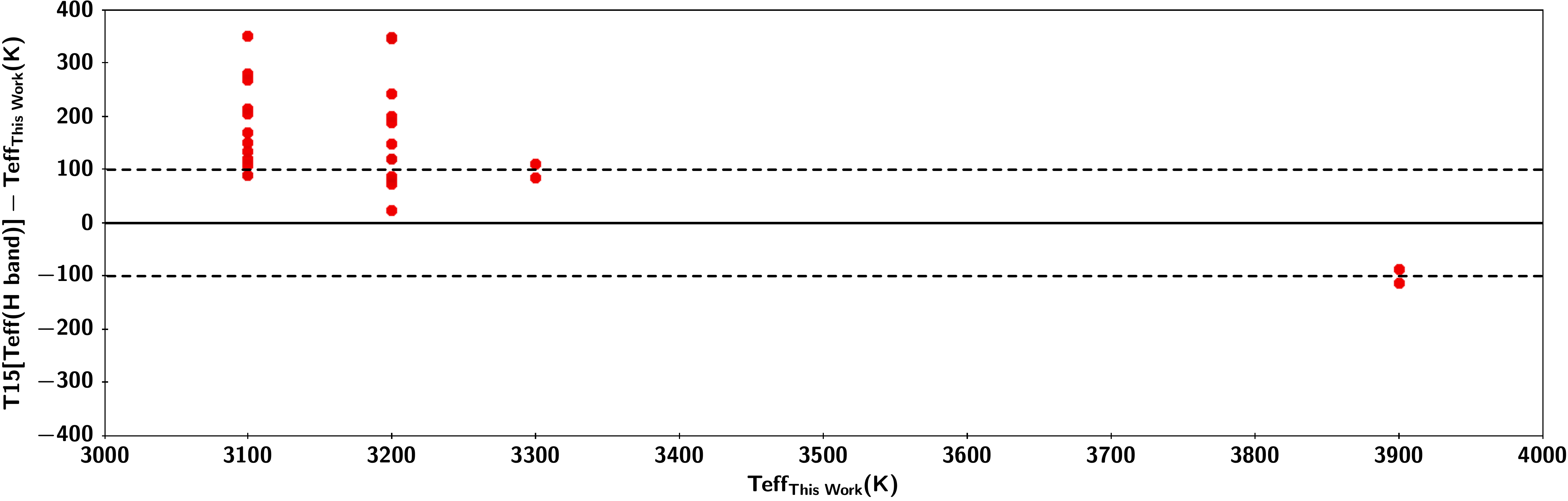}
        
        \hspace{0mm}
        
        \includegraphics[width=8.0cm,height=4.0cm]{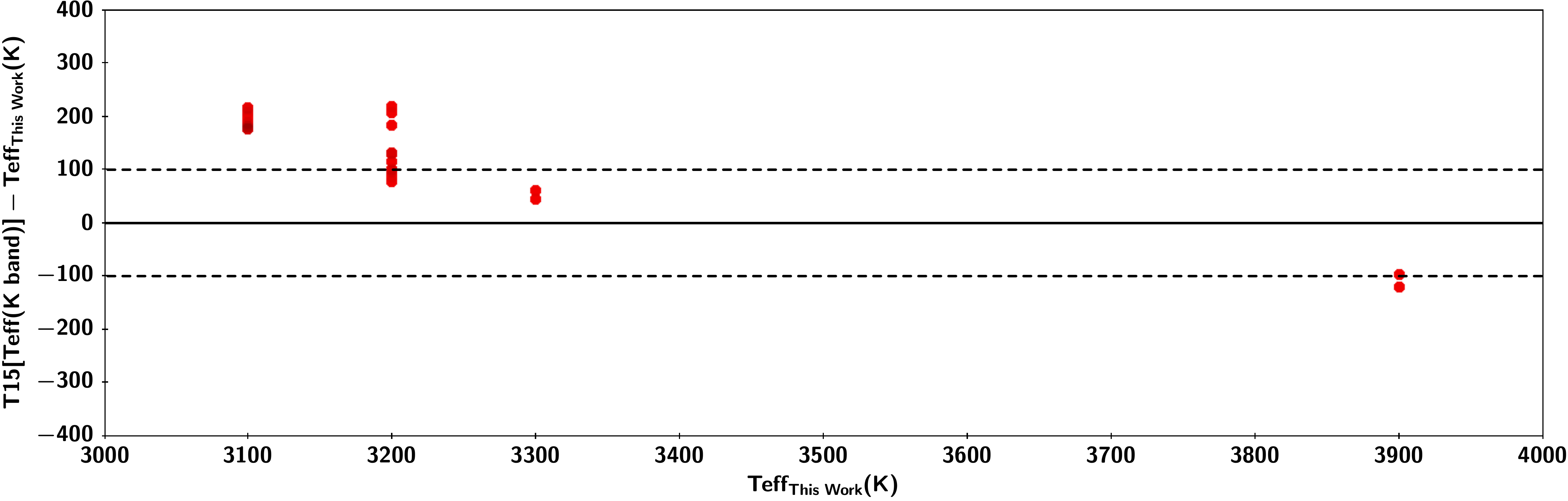}
        
        \caption{Difference between the $\teff$ calibrations from \citet[][T15]{Terrien2015}, estimated for the M dwarfs from \cite{Mann2013b} J (top), H (middle) and K$_s$ (bottom) calibrations and $\teff$ from this work. On the horizontal axis we show the $\teff$  that we infer from our best fit BT-Settl model used in this work. The black full line represents the origin and the dashed black lines represent the error from the grid size of 100K.}
        \label{Fig6}
\end{figure}

\begin{figure}[!thbp]
        \centering
          \includegraphics[width=8.0cm,height=4.0cm]{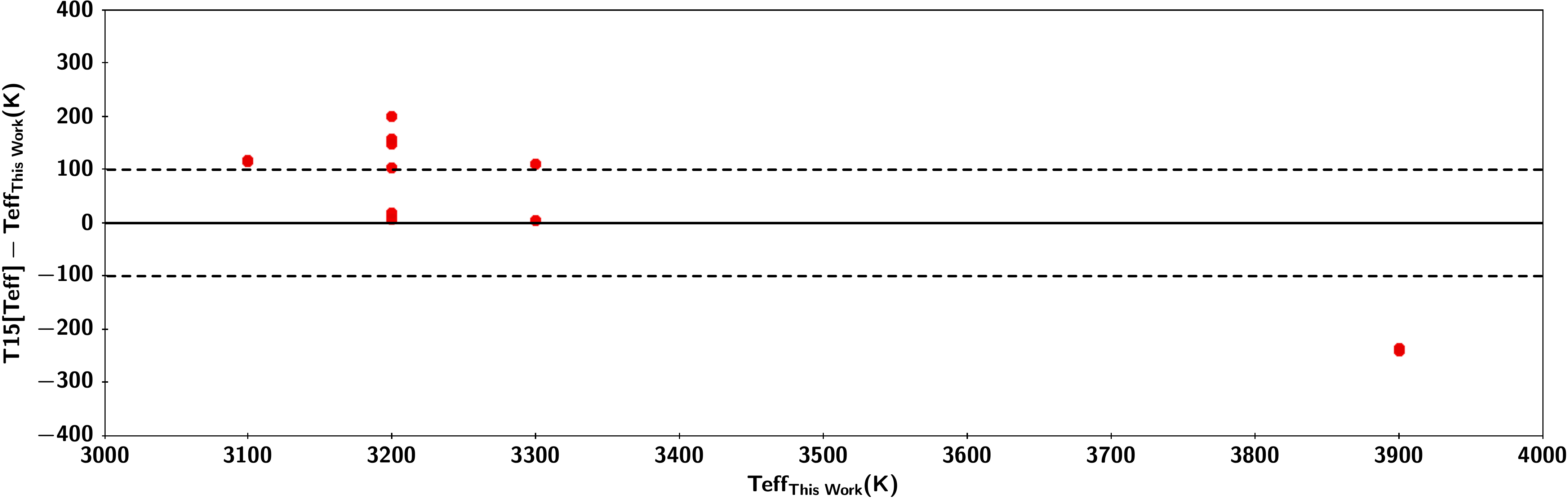}
        \caption{Difference between the $\teff$ calibrations from \citet[][T15]{Terrien2015} , estimated for the M dwarfs from H-band
relationships given by \cite{Newton2015}, calibrations, and $\teff$ from this work. On the horizontal axis we show the $\teff$ that we infer
from our best fit BT-Settl model used in this work. The black full line represents the origin and the dashed black lines represent the error from the
grid size of 100K.}
        \label{Fig7}
\end{figure}


\begin{table*}[!thbp]

        \caption{Stellar parameters of the observed targets determined by minimizing $\chi^2$ . The uncertainty in $\teff$ (K) is $\pm$ 100 K, whereas for $\mathrm{log}\,g~$(cm~$s^{-2}$) and [Fe/H] is given below.}
        \begin{tabular}{cccc}
                \hline
                2MASS ID&This study&\cite{Terrien2015} &\cite{Terrien2015} \\
                2MASS J&$\teff$/ $\logg$ /[Fe/H]&$\teff$,~[Fe/H]& $\teff$,~[Fe/H]\\
                        &&using \cite{Mann2013b} &using \cite{Newton2014}\\
                        && J, H and K calibration&calibration\\
                        \hline\\     
00131578+6919372        &3200 / 5.5$\pm$0.3 / -0.3$\pm$0.04&--&--\\
00321574+5429027  &3200 / 5.5$\pm$0.3 / -0.2$\pm$0.04&3366/3271/3285, -0.03/-0.08/-0.05&3206/+0.00 \\
00350487+5953079  &3100 / 5.5$\pm$0.3 / -0.0$\pm$0.05&--&--\\
01195227+8409327  &3100 / 5.5$\pm$0.3 / -0.3$\pm$0.06&--&--\\
02085359+4926565  &3200 / 5.5$\pm$0.3 / -0.1$\pm$0.05&3280/3285/3330, +0.08/+0.03/+0.05&3347/+0.14\\
03152943+5751330  &3200 / 5.5$\pm$0.3 / -0.3$\pm$0.05&--&--\\
3305473+7041145         &3200 / 5.5$\pm$0.3 / -0.3$\pm$0.05&--&--\\
03425325+2326495  &3200 / 5.5$\pm$0.3 / -0.0$\pm$0.05&--&--\\
4063732+7916012    &3100 / 5.5$\pm$0.2 / -0.0$\pm$0.06&--&--\\
04125880+5236421  &3100 / 5.5$\pm$0.3 / -0.0$\pm$0.05&3026/3304/3276, +0.02/-0.08/-0.02&--/+0.06\\
05011802+2237015  &3200 / 5.5$\pm$0.5 / -0.5$\pm$0.04&--/3223/3277, +0.21/+0.03/+0.12&--/+0.20\\
05030563+2122362  &3100 / 5.5$\pm$0.2 / -0.1$\pm$0.07&--/3223/3277, +0.13/+0.02/+0.02&--/+0.05\\
05210188+3425119  &3100 / 5.5$\pm$0.3 / -0.5$\pm$0.04&--&--\\
05470907-0512106   &3100 / 5.5$\pm$0.3 / -0.3$\pm$0.06&--&--\\
06115599+3325505  &3100 / 5.5$\pm$0.2 / -0.1$\pm$0.07&3099/3207/3276, +0.02/+0.12/+0.01&--/+0.13\\
06320207+3431132  &3200 / 5.5$\pm$0.3 / -0.4$\pm$0.05&3126/3388/3313, -0.03/-0.05/-0.09&--/+0.03\\
07140394+3702459  &3000 / 5.5$\pm$0.2 / -0.5$\pm$0.11&--&--\\
08501918+1056436  &3100 / 5.5$\pm$0.2 / -0.0$\pm$0.06&--&--\\
09301445+2630250  &3300 / 5.0$\pm$0.5 / -0.3$\pm$0.05&3285/3384/3359, +0.04/+0.13/+0.13&3410/+0.21\\
10162955+0318375  &3200 / 5.5$\pm$0.3 / -0.2$\pm$0.03&3345/3399/3328, -0.03/+0.03/-0.05&3217/+0.00\\
11005043+1204108  &3100 / 5.5$\pm$0.2 / -0.5$\pm$0.11&--/3304/3276, +0.12/-0.07/-0.11&--/+0.05\\
11054316+1014093  &3200 / 5.0$\pm$0.5 / -0.0$\pm$0.05&3422/3547/3418, -0.08/-0.10/-0.05&3357/+0.01\\
11091225-0436249   &3900 / 4.5$\pm$0.5 / -0.3$\pm$0.04&3670/3786/3803, -0.04/-0.14/-0.04&3659/-0.07\\
11474074+0015201  &3200 / 5.5$\pm$0.3 / -0.4$\pm$0.04&--/3320/3300, +0.10/+0.1/5 +0.03&--/+0.17\\
12045611+1728119  &3200 / 5.5$\pm$0.2 / -0.1$\pm$0.07&3235/3318/3330, -0.09/-0.11/-0.03&3303/+0.05\\
12232063+2529441  &3300 / 5.0$\pm$0.5 / -0.4$\pm$0.04&3267/3409/3344, -0.05/-0.01/+0.05&3303/+0.05\\
12265737+2700536  &3100 / 5.5$\pm$0.3 / -0.0$\pm$0.06&--/3304/3280, +0.13/-0.05/ +0.02&--/+0.11\\
13085059+1622039  &3200 / 5.5$\pm$0.4 / -0.4$\pm$0.04&3533/3545/3407, -0.15/-0.10/-0.15&--/+0.11\\
13345147+3746195  &3200 / 5.5$\pm$0.3 / -0.1$\pm$0.04&3219/3348/3297, +0.13/-0.01/+0.14&--/+0.22\\
13451104+2852012  &3200 / 5.0$\pm$0.5 / -0.4$\pm$0.04&3385/3441/3383, -0.09/-0.16/-0.09&3399/-0.11\\
14592508+3618321  &3200 / 5.5$\pm$0.3 / -0.0$\pm$0.04&--&--\\
16370146+3535456  &3100 / 5.5$\pm$0.2 / -0.5$\pm$0.04&--&--\\
18451027+0620158  &3900 / 4.5$\pm$0.5 / -0.4$\pm$0.04&3707/3812/3779, +0.03/+0.07/-0.03&3664/-0.05\\
18523373+4538317  &3100 / 5.5$\pm$0.2 / -0.0$\pm$0.07&3169/3219/3285, -0.00/-0.08/-0.06&--/-0.03\\
18562628+4622532  &3100 / 5.5$\pm$0.3 / -0.0$\pm$0.05&3091/3379/3307, +0.06/-0.14/-0.03&--/+0.05\\
19051739+4507161  &3100 / 5.5$\pm$0.3 / -0.2$\pm$0.04&3339/3314/3313, -0.06/-0.23/-0.17&3215/-0.14\\
19071270+4416070  &3100 / 5.5$\pm$0.3 / -0.3$\pm$0.06&3163/3269/3288, +0.2/2-0.0/2+0.19&--/+0.25\\
9081576+2635054    &3100 / 5.5$\pm$0.3 / -0.4$\pm$0.06&4747/3449/3280, +0.77/+0.10/+0.26&--/+0.29\\
19084251+2733453  &3100 / 5.5$\pm$0.3 / -0.2$\pm$0.04&--/3368/3316,+0.42/-0.28/-0.30&3217/-0.32\\
19321796+4747027  &3100 / 5.5$\pm$0.3 / -0.3$\pm$0.05&--&\\
19333940+3931372  &3200 / 5.5$\pm$0.3 / -0.1$\pm$0.05&--&\\
19430726+4518089  &3100 / 5.5$\pm$0.2 / -0.5$\pm$0.06&--&\\
19443810+4720294  &3100 / 5.5$\pm$0.3 / -0.5$\pm$0.04&--&\\
19510930+4628598  &3200 / 5.5$\pm$0.3 / -0.0$\pm$0.07&--/3279/3295, +0.06/-0.09/-0.07&--/+0.07\\
21105881+4657325  &3300 / 5.0$\pm$0.5 / -0.2$\pm$0.06&--&\\
                \hline\\
                \label{Table2}
        \end{tabular}
\end{table*}

\begin{figure*}[!thbp]
        \centering
        \includegraphics[width=15.5cm,height=15cm,angle=-90]{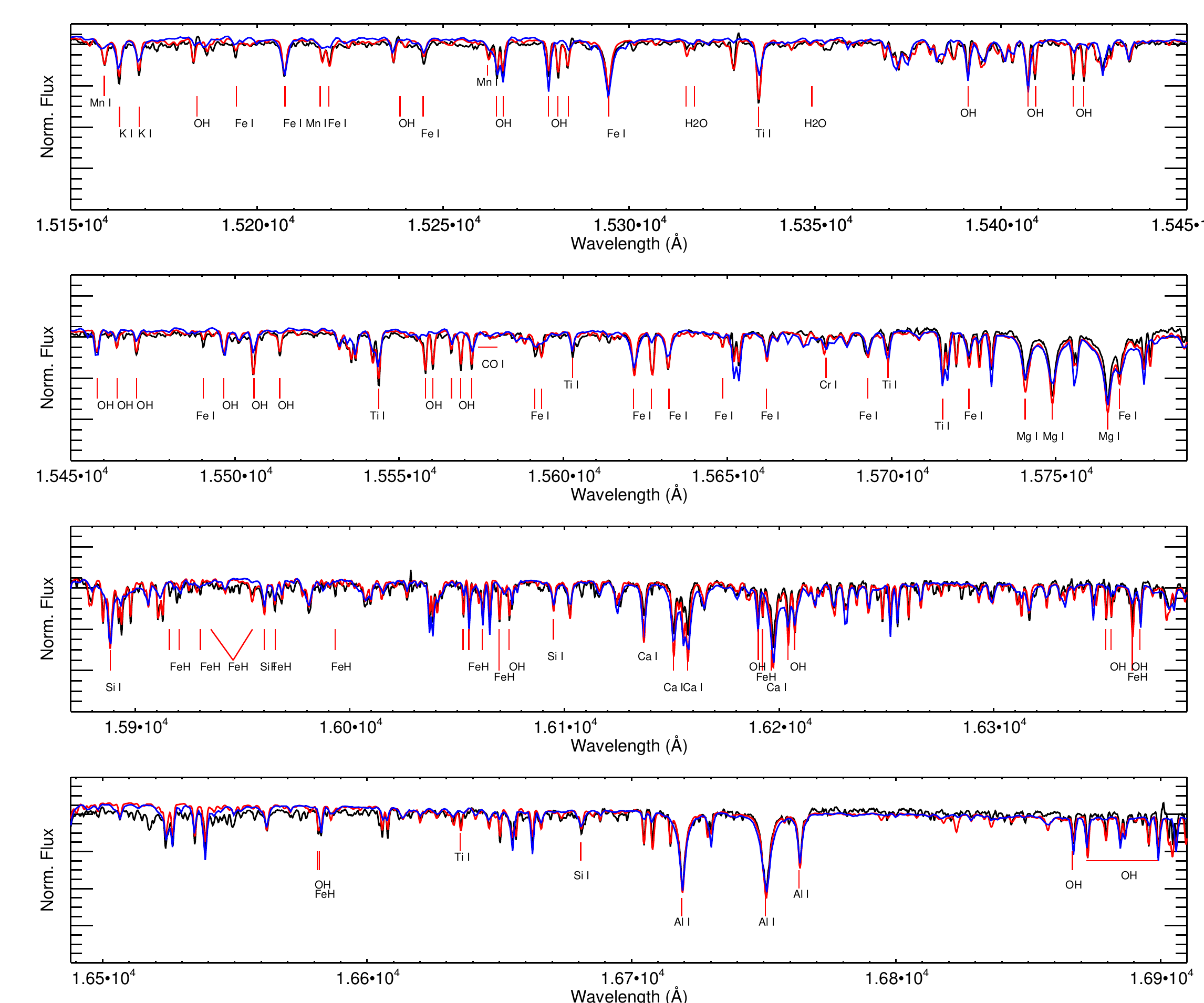}
\caption{APOGEE spectra of 2M11091225-0436249 (black) of spectral type M0.5 is compared with the best-fit BT-Settl (red) and MARCS model (blue). The best fit value for $\teff$, $\logg$ and [Fe/H] is 3900/4.5/-0.3.}
\label{Fig8}
\end{figure*}

\begin{figure*}[!thbp]
        \centering
        \includegraphics[width=15.5cm,height=15cm,angle=-90]{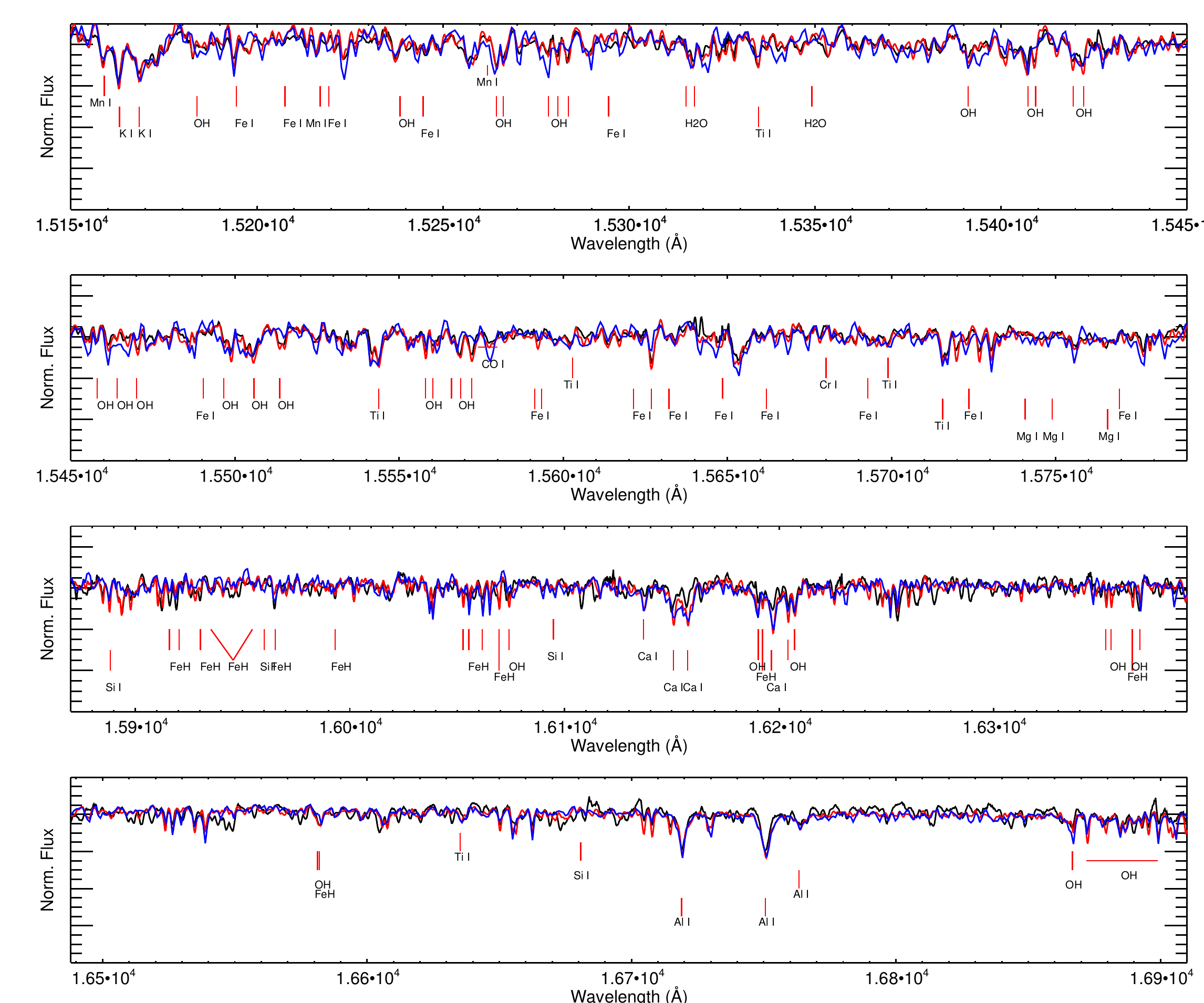}
\caption{APOGEE spectra of 2M08501918+1056436 (black) of spectral type M5.0 is compared with the best-fit BT-Settl (red) and MARCS model (blue). The best fit value for $\teff$, $\logg$ and [Fe/H] is 3100/5.5/-0.0.}
\label{Fig9}
\end{figure*}

\section{Summary and Discussion}
\label{discussion}

The high-resolution spectra with good signal-to-noise ratio of M dwarfs is very important and necessary to determine the $\teff$, $\logg,$ and [Fe/H], and also the individual element abundances to a high accuracy. This paper presents the results from the spectral synthesis analysis to determine the fundamental parameters form the high-resolution APOGEE H-band spectra for early to mid M dwarfs using the updated BT-Settl model. In the NIR, particularly in H band, the BT-Settl model has never been tested before with the high-resolution spectra of M dwarfs. Therefore, our present study constitutes a benchmark for model atmospheres of low-mass stars in NIR. The physical parameters $\teff$, $\logg,$ and [Fe/H] for stars of our sample is determined by comparing the observed spectra with the synthetic spectra. The main purpose of this paper is to disentangle the parameter space ($\teff$, $\logg,$ and [Fe/H]) with independent information on atmospheric parameters. The comparison of observed spectra with the synthetic spectra is crucial to reveal the inaccuracy or incompleteness of the opacities used in the model. The atmospheric parameters derived from the comparison between our sample and the BT-Settl model are summarized in Table \ref{Table2}. For example, Figs.~\ref{Fig4} and \ref{Fig5} show the comparison of the best-fit BT-Settl model (red) with the star of spectral type M1.0 and M3.0 (black) in our sample. Their best fit parameters are given in Table \ref{Table2}. The specific strengths of the CO, OH, and FeH-band heads are very well reproduced by the synthetic spectra over the entire M dwarf sequence, showing that the nosy pattern visible at this high spectral resolution is not noise. 

The BT-Settl models also predict and reproduce the shape of various atomic lines such as Ca I, Na I, K I, Si I, Mg I, Al I, Ti I rather well and their strengths are well fitted. The observed atomic lines in the spectra are broader and shallower than those predicted by the BT-Settl model in the cooler M dwarfs (spectral type M3 or later). The qualitative behavior of the K I, Al I, Mg I, Ti I and Ca I lines is well reproduced by the BT Settl model as compared to the strong pressure-broadening wings in the early to mid M dwarfs. In the early M dwarfs, the cores of the observed K I, Al I, Mg I and Ca I lines are still visible. The broader absorption component of the atomic lines becomes saturated in M dwarfs later than M6 which were extending a few tens to one hundred angstroms. 

The $\teff$ is the parameter that causes the largest uncertainty while determining the other stellar parameters of M dwarfs; their metallicity in particular. Our results for $\teff$ are in good agreement with the $\teff$ as a function of spectral type given in \cite{Rajpurohit2013}. Now we compare our $\teff$, $\logg,$ and [Fe/H] determination to other works such as \cite{Terrien2015,Schmidt2016}.  \cite{Terrien2015} measured the $\teff$ for the M dwarfs using color--$\teff$ relations with the method described by \cite{Mann2013b} along with different temperature indices such as H$_2$O-K2 \citep{Roja2012}, H$_2$O-H \cite{Terrien2012} and \cite{Mann2013}. Figure ~\ref{Fig6} shows the comparison of our measured $\teff$ with \cite{Terrien2015} which clearly shows that \cite{Terrien2015} overestimates in lower $\teff$ and underestimates in higher $\teff$ among the various calibrations using J, H and K$_s$ bands, when compared to our $\teff$ determinations. This discrepancy could be due to that fact that their determination was based on near-infrared spectra using the SpeX spectrograph which has significantly lower resolution, and many of their individual determinations were from the J/H/K bands which give relatively inconsistent results. These empirical relations give smaller errors as compared to NIR but they are not as precise as model-fitting techniques. We have also compared the $\teff$ calculated by \cite{Terrien2015} based on H-band atomic feature strengths such as Al I, Mg I, K I, Si I (Fig ~\ref{Fig7}) using the strength of atomic features studied in \cite{Newton2015}. We find an offset of around 200 K between our $\teff$ and \cite{Terrien2015} which could be due to the fact that \cite{Newton2015} used a limited number of atomic lines for equivalent width in their analysis where the accurate continuum placement could be the issue.

In four of the stars common common to both ours and the \cite{Schmidt2016} sample, we find that for stars 2MASSJ 11091225-0436249 and 2MASSJ 18451027+0620158, the $\teff$ by \cite{Schmidt2016} is 200 to 300K lower than our measurements, whereas for 2MASSJ 19333940+3931372 and 2MASSJ 21105881+4657325 the $\teff$ by \cite{Schmidt2016} is 200 to 300K higher. \cite{Schmidt2016} determine the $\teff$, $\logg,$ and [Fe/H] of late-K and early-M dwarfs selected from the APOGEE spectroscopic survey using ASPCAP (APOGEE Stellar Parameters and Chemical Abundances Pipeline) \citep{Garcia2016}. ASPCAP uses APOGEE ATLAS9 models \citep{Meszaros2012}. For this same set of four targets, we have compared $\logg$ and [Fe/H] with \cite{Schmidt2016} and found a systematic offset of around ~0.5 dex to 1.0 dex.  We have also compared the best-fit BT-Settl model (red) and MARCS model (blue) with observed spectra of 2M11091225-0436249 and 2M08501918+1056436 (back). We have chosen the identical atmospheric parameters for MARCS model as mentioned in Table \ref{Table2}. We obtain the MARCS \citep{Gustafsson2008} model which was calculated in 2012 and distributed on the MARCS website\footnote{http://marcs.astro.uu.se}. It is clear from Figs.~\ref{Fig8} and \ref{Fig9} that in the MARCS model, many OH, CO, and FeH bands are missing. Also, the line strength of various atomic species, such as K I, Ti I, Ca I and Al I, is weaker in the MARCS model than in the BT-Settl model which could be due to low-resolution flux samples as provided on the MARCS website. This discrepancy may also be due to the use of somewhat different assumptions concerning convection, and input data such as continuous opacities in MARCS and ATLAS9 models as compared to the BT-Settl model. Exploring such effects is beyond the scope of this study but a proper way would be to compare the best fit parameters derived using different sets of models, which would provide information on model systematics. For the BT-Settl model, a crucial test is to check its consistent accuracy.

\begin{figure}[!thbp]
        \centering
        \includegraphics[width=8.0cm,height=4.0cm]{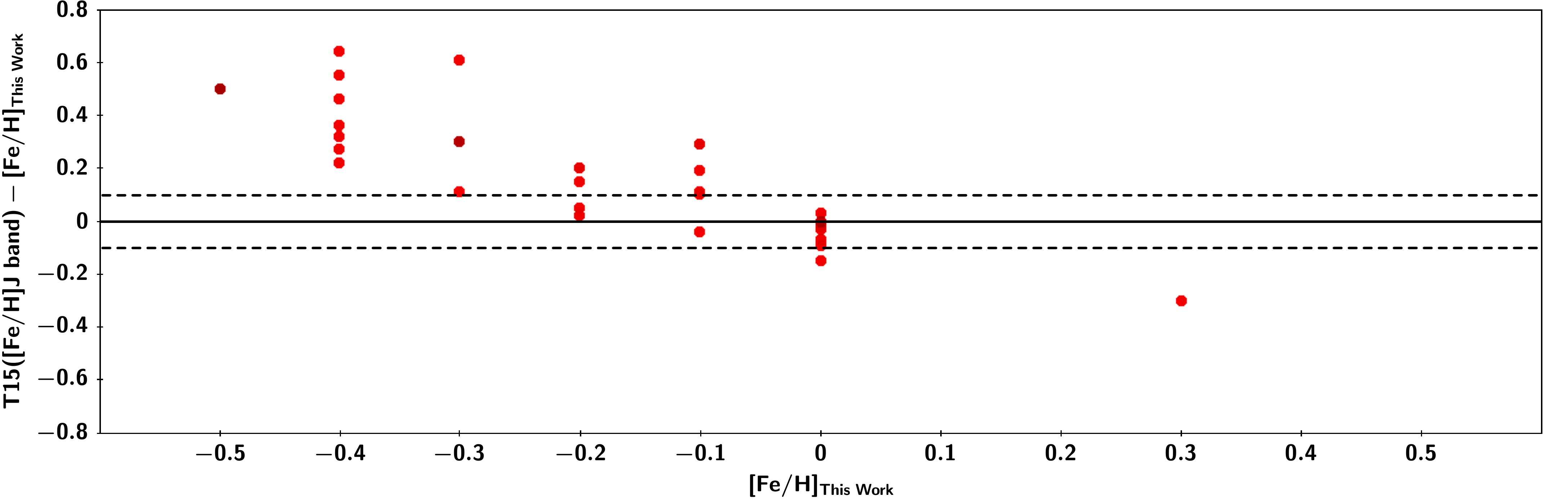}
        \caption{Difference between the [Fe/H] calibrations from \citet[][T15]{Terrien2015}, estimated for the M dwarfs from H-band relationships given by \cite{Newton2014} calibrations and [Fe/H] from this work. On the horizontal axis we show the [Fe/H] that we infer from our best fit BT-Settl model used in this work. The black full line represents the origin and the dashed black lines represent the error from the
grid size of 0.1 dex.}
        \label{Fig10}
\end{figure}

\begin{figure}[!thbp]
        \centering
                \includegraphics[width=8.0cm,height=4.0cm]{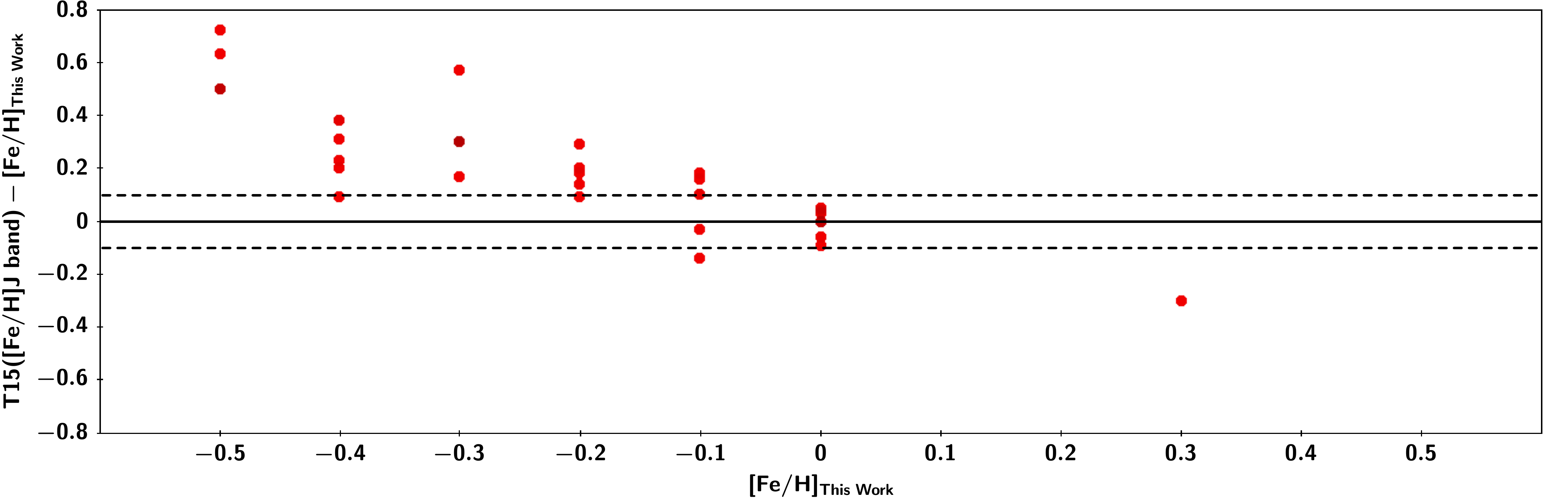}
                \hspace{0mm}
        
          \includegraphics[width=8.0cm,height=4.0cm]{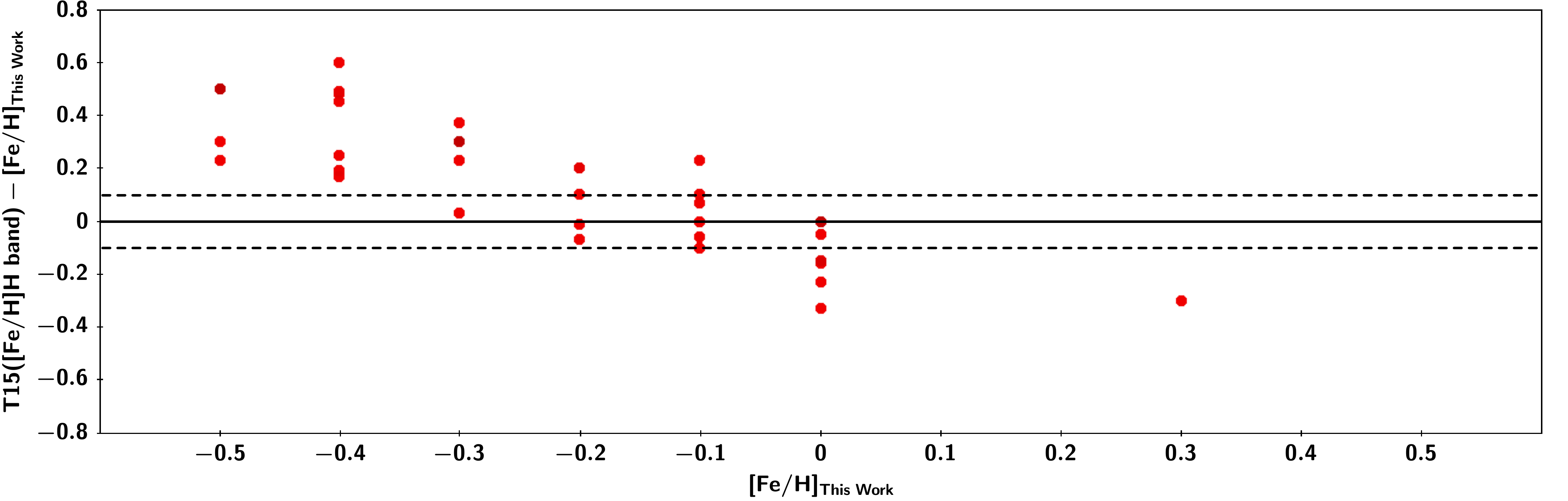}

        \hspace{0mm}
        
        \includegraphics[width=8.0cm,height=4.0cm]{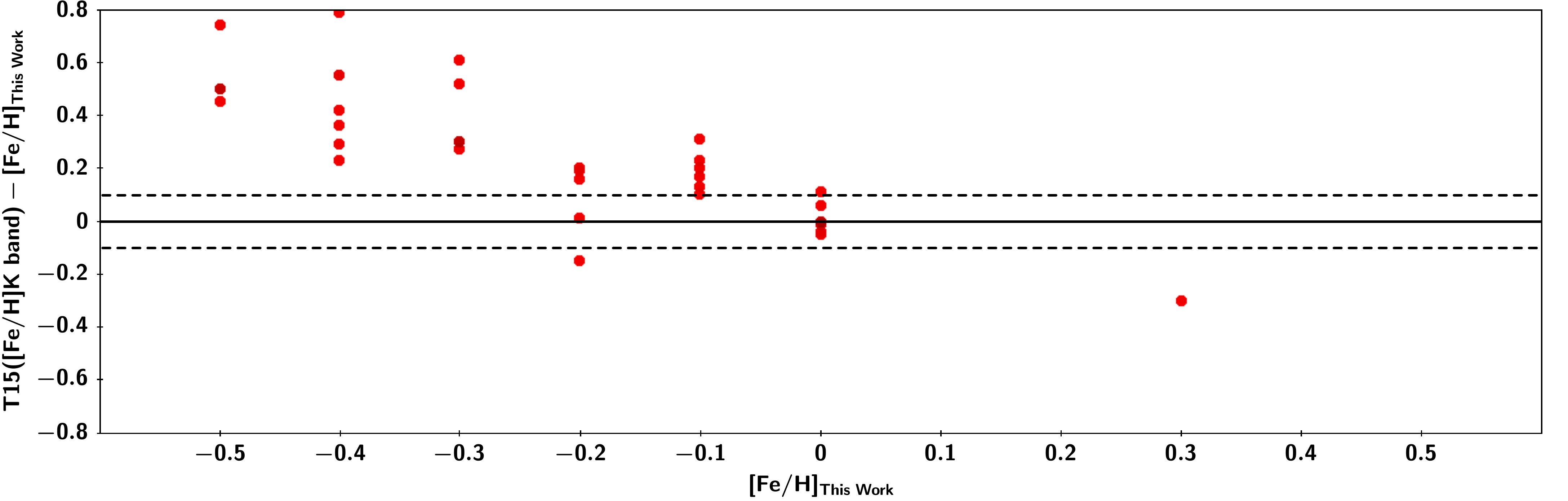}

        \caption{Difference between the [Fe/H] calibrations from \citet[][T15]{Terrien2015}, estimated for the M dwarfs from \cite{Mann2013b} J (top), H (middle) and K$_s$ (bottom) calibrations and [Fe/H] from this work. On the horizontal axis we show the [Fe/H] that we infer from our best fit BT-Settl model used in this work. The black full line represents the origin and the dashed black lines represent the error from the grid size of 0.1 dex.}
        \label{Fig11}
\end{figure}

Metallicity is a parameter which cannot be constrained independently, but can be determined from spectroscopic analysis. We have also compared our [Fe/H] determination with spectroscopically determined metallicity estimates from \cite{Terrien2015} (see Figs.~\ref{Fig10} and~\ref{Fig11}). \cite{Terrien2015} used both J, H, and K$_s$ band calibration given by \cite{Mann2013b} and the combinations of EW that effectively trace stellar metallicity from the H-band spectra given by \cite{Newton2014}. \cite{Terrien2015} estimated the metallicities of the M dwarfs using the  EW of the NaI feature at 2.2 $\mu$m in the K band of Infrared Telescope Facility (IRTF) spectra as used by \cite{Newton2014}. We  find an average deviation of 0.2 to 0.4 dex in [Fe/H] from \cite{Newton2014} and \cite{Terrien2015}. A possible explanation for this deviation could problems or differences in the determination of $\teff$. The [Fe/H] and $\teff$ are dependent to the point where there is normally a degeneracy of models based on this interdependence. The different parameter combinations of $\teff$ and $\logg$ can produce the same [Fe/H] at low resolution. These deviations could also be due to the fact that the BT-Settl model provides a better description of the M dwarf atmospheres and therefore more accurate metallicities can be derived than when using other methods, which is also pointed out by \cite{Lindgren2017}.

The recent improvement in the BT-Settl model atmosphere could have implications beyond those noted in this study. The description of various physical process at these low temperatures is well explained by BT-Settl models. These models now provide a better fit to the high-resolution spectroscopic observations of M dwarfs and help in accurately determining their atmospheric parameters. To address our offset in metallicity using different sets of model atmospheres, we also made a comparison study with the MARCS model. This comparison suggests that the BT-Settl models describe cool atmospheres more accurately than the MARCS model. We plan to use our method along with the grid of these new BT-Settl models to estimate the stellar parameter of M dwarfs both in optical and in NIR spectra and photometry simultaneously to minimize the differences. The improvements in BT-Settl \citep{Allard2013}, achieved with the revision of solar abundances
by \cite{Asplund2009} and \cite{Caffau2011}, and by including updated atomic
and molecular line opacities that dominate both in the optical and NIR
range M dwarfs, greatly help to reproduce the extensive and complex molecular absorption band systems present in M dwarf atmospheres.

As compared to other models, the line strength and shape of various atomic and molecular features is very well reproduced by the BT-Settl models but there is still need for improvement in the regions where the fit is not good. This could be due to the lack of various opacity line lists in the model; in particular the FeH line list is missing in the H bandpass. Currently, the ExoMol group is developing an accurate and complete line list for TiO which is the next step to including them in the BT-Settl model before computing detailed model atmosphere grids and interior and evolution models at finer steps in the atmospheric parameters. The three-dimensional radiative hydrodynamics simulations and radiative transfer will help in understanding the effects of temperature inhomogeneities in the atmosphere which begin to have greater impact on the spectrum formation.

\begin{acknowledgements}
The research leading to these results has received funding from the French "Programme National de Physique Stellaire" and the Programme National de Planetologie of CNRS (INSU). The computations were performed at the {\sl P\^ole Scientifique de Mod\'elisation Num\'erique} (PSMN) at the {\sl \'Ecole Normale Sup\'erieure} (ENS) in Lyon, and at the {\sl Gesellschaft f{\"u}r Wissenschaftliche Datenverarbeitung G{\"o}ttingen} in collaboration with the Institut f{\"u}r Astrophysik G{\"o}ttingen. DH is supported by Sonderforschungsbereich SFB 881 ``The Milky Way System'' (subproject A4) of the German Research Foundation (DFG). G. D. C. Teixeira acknowledges support from the fellowship PD/BD/113478/2015 funded by FCT (Portugal) and POPH/FSE (EC). This work was supported in part by Funda\c{c}ão para a Ci\^encia e a Tecnologia (FCT) through national funds (UID/FIS/04434/2013) and by FEDER through COMPETE2020 (POCI-01-0145-FEDER-007672). We also want to thank the anonymous referee for useful comments, which helped improve the paper.\\

Funding for the Sloan Digital Sky Survey IV has been provided  by  the  Alfred  P.  Sloan  Foundation,  the  U.S. Department of Energy Office of Science, and the Participating  Institutions.   SDSS-  IV  acknowledges  support and  resources  from  the  Center  for  High-Performance Computing at the University of Utah.  The SDSS web site is www.sdss.org.  SDSS-IV is managed by the Astrophysical  Research  Consortium  for  the  Participating Institutions  of  the  SDSS  Collaboration  including  the Brazilian Participation Group, the Carnegie Institution for  Science,  Carnegie  Mellon  University,  the  Chilean Participation  Group,  the  French  Participation  Group, Harvard-Smithsonian Center for Astrophysics, Instituto de Astrof\'{?}sica de Canarias, The Johns Hopkins University, Kavli Institute for the Physics and Mathematics of the Universe (IPMU) / University of Tokyo, Lawrence Berkeley National Laboratory, Leibniz Institut f{\"u}r Astrophysik Potsdam (AIP), Max-Planck-Institut f{\"u}r Astronomie  (MPIA  Heidelberg),  Max-Planck-Institut  f{\"u}r Astrophysik (MPA Garching), Max-Planck-Institut f{\"u}rExtraterrestrische  Physik  (MPE),  National  Astronomical  Observatory  of  China,  New  Mexico  State  University,  New  York  University,  University  of  Notre  Dame, Observat\'{o}rio Nacional / MCTI, The Ohio State University, Pennsylvania State University, Shanghai Astronomical Observatory, United Kingdom Participation Group, Universidad Nacional Aut\'{o}noma de M\'{e}xico, University
of Arizona, University of Colorado Boulder, University of Oxford, University of Portsmouth, University of Utah, University of Virginia,  University of Washington,  University  of  Wisconsin,  Vanderbilt  University,  and  Yale University.
\end{acknowledgements}

\bibliographystyle{aa}
\bibliography{ref}
\end{document}